\documentclass{jpsj2}
\usepackage{graphicx,amsmath,bm}

\def\eq{\hspace{-3mm}&=&\hspace{-3mm}}
\def\espace{\hspace{-3mm}&&\hspace{-3mm}}
\def\eequiv{\hspace{-3mm}&\equiv&\hspace{-3mm}}

\def\d{\mbox{\rm d}}
\def\e{\,\mbox{\rm e}}
\def\tr{\mbox{\rm tr}\,}
\def\det{\mbox{\rm det}\,}
\def\im{\mbox{\rm i}}
\newcommand{\vecvar}[1]{\mbox{\boldmath$#1$}}
\def\p{\partial}
\def\smalli{\mbox{\scriptsize \rm i}}
\def\k{\kappa}

\makeatletter
\def\@citess#1{\textsuperscript{#1)}}
\title{Matter-Wave Bright Solitons with a Finite Background in Spinor Bose-Einstein Condensates}

\author{Tetsuo \textsc{Kurosaki}\thanks{E-mail address: kurosaki@monet.phys.s.u-tokyo.ac.jp} and Miki \textsc{Wadati}}

\inst{Department of Physics, Graduate School of Science, University of Tokyo, Tokyo 113-0033}

\abst{We investigate dynamical properties of bright solitons with a finite background in the $F=1$ spinor Bose-Einstein condensate (BEC), based on an integrable spinor model which is equivalent to the matrix nonlinear Schr\"{o}dinger equation with a self-focusing nonlineality. We apply the inverse scattering method formulated for nonvanishing boundary conditions. The resulting soliton solutions can be regarded as a generalization of those under vanishing boundary conditions. One-soliton solutions are derived in an explicit manner. According to the behaviors at the infinity, they are classified into two kinds, domain-wall (DW) type and phase-shift (PS) type. The DW-type implies the ferromagnetic state with nonzero total spin and the PS-type implies the polar state, where the total spin amounts to zero. We also discuss two-soliton collisions. In particular, the spin-mixing phenomenon is confirmed in a collision involving the DW-type. The results are consistent with those of the previous studies for bright solitons under vanishing boundary conditions and dark solitons. As a result, we establish the robustness and the usefulness of the multiple matter-wave solitons in the spinor BECs.}

\kword{spinor Bose-Einstein condensate, internal degrees of freedom, matrix nonlinear Schr\"{o}dinger equation, integrable model, bright soliton, nonvanishing boundary condition, multi-component Gross-Pitaevskii equation, spin exchange coupling, spin switching}

\begin{document}
\maketitle

\newpage
\section{Introduction} %% No sections necessary for express letters, letters and short notes

In 2002, matter-wave bright solitons in quasi-one-dimensional (1D) Bose-Einstein condensates (BECs) were observed experimentally \cite{Brightsol1, Brightsol2}. Bright solitons propagate in most cases with much larger amplitudes than dark solitons \cite{Darksol1, Darksol2}, and are expected to have the potential for various applications such as coherent transport and atom interferometry. Soliton propagation in BEC can be described by the Gross-Pitaevskii (GP) equation. The GP equation, called the nonlinear Schr\"{o}dinger (NLS) equation in nonlinear science, is integrable and has soliton solutions in a one-dimensional and uniform system. Recent experimental and theoretical advances about matter-wave bright solitons are reviewed, for instance, in ref. \citen{Sol_rev}.

The experimental creation of matter-wave solitons has been so far achieved only for single-component BEC. It is, nevertheless, very interesting to consider soliton propagation in BEC with internal degrees of freedom, so-called, spinor BEC. When BEC of ultracold alkali atoms is trapped exclusively by optical means, the hyperfine spin of atoms remains liberated. The spinor BEC was realized in such a way \cite{spinor_BEC_MIT1, spinor_BEC_MIT2, spinor_BEC_MIT3}. Internal degrees of freedom endow solitons with a multiplicity. The multiple solitons will show a rich variety of dynamics. Here, we focus on the boson system in the $F=1$ hyperfine spin state, exemplified by $^{23}$Na, $^{39}$K and $^{87}$Rb. The multi-component GP equation for $F=1$ spinor BEC turns to an integrable model at special points, which is mathematically equivalent to the matrix NLS equation. An integrable model with a self-focusing nonlineality enables one to perform exact analysis via the inverse scattering method (ISM) for the matrix NLS equation \cite{Tsuchida}. In particular, bright soliton solutions under \textit{vanishing boundary conditions} (VBC) are obtained, whose properties are investigated in refs. \citen{IMW1} and \citen{IMW2}. Recently, the ISM for the matrix NLS equation under \textit{nonvanishing boundary conditions} (NVBC) is formulated \cite{ISM_NVBC}. Dark solitons in the $F=1$ spinor BEC can be investigated by applying the ISM under NVBC to an integrable model with a self-defocusing nonlineality \cite{UIW}. Although the ISM under NVBC is dedicated mainly to the self-defocusing case, we note that this technique is also applicable to an integrable model with a self-focusing nonlineality, which makes us available to bright soliton solutions with a finite background.

In this paper, the detail of matter-wave bright solitons in the quasi-1D $F=1$ spinor BEC is further investigated, based on an integrable model. We consider matter-wave spinor bright solitons traveling on a finite background of the condensate. We write down explicitly new soliton solutions, and verify that the obtained soliton solutions have the similar properties compared to those without a background. In the usual experimental setups, the condensates are confined in a finite-size regime, and the matter-wave bright solitons will accompany a finite background. The study given in this paper is meaningful in such realistic circumstances. 

The paper is organized as follows. In $\S$ \ref{sec:spinorGP}, the GP equation for quasi-1D $F=1$ spinor BEC is introduced. In particular, the integrable model is presented. There, the interactions between two atoms are supposed to be inter-atomic attractive and ferromagnetic, which lead to bright solitons. In $\S$ \ref{sec:bright soliton under NVBC}, the inverse scattering method under nonvanishing boundary conditions is applied to the integrable model. This application leads to bright soliton solutions with a finite background. Several conserved quantities of the model are also provided. One-soliton solutions are investigated in $\S$ \ref{sec:one-soliton}. The spin states of one-solitons are classified, assuming that discrete eigenvalues are purely imaginary. Two-soliton solutions are discussed in $\S$ \ref{sec:two-soliton}. The last section, $\S$ \ref{sec:conclusion}, is devoted to the concluding remarks.     

\section{GP Equation for $F=1$ Spinor Bose-Einstein Condensates}
\label{sec:spinorGP}

For BEC of ultracold alkali atoms, the mean-field theory works well, because almost all atoms go into condensation and the condensate is dilute. In this paper, we deal with the  quasi-one-dimensional system. Atoms in the $F=1$ hyperfine spin state have three magnetic substates labeled by the magnetic quantum number $m_F=1,0,-1$. The system is characterized by a vectorial field operator with the components corresponding to each substate, $\hat{\vecvar{\Phi}}=(\hat{\Phi}_1,\hat{\Phi}_0,\hat{\Phi}_{-1})^T$, satisfying equal-time commutation relations:
\begin{equation}
[\hat{\Psi}_{\alpha}(x,t), \hat{\Psi}^\dagger_{\beta}(x',t)]=\delta_{\alpha\beta}\delta (x-x'),
\end{equation}
where the subscripts $\alpha,\beta$ take on $1,0,-1$. In the framework of the mean-field theory for BEC, the quantum field is replaced with the order parameter:
\begin{equation}
\vecvar{\Phi}(x,t)\equiv \langle \hat{\vecvar{\Phi}}(x,t) \rangle=\left(\Phi_1(x,t),\Phi_0(x,t),\Phi_{-1}(x,t)\right)^T.
\end{equation}
$\vecvar{\Phi}(x,t)$ is often called the spinor condensate wavefunction, which is normalized to the total number of atoms $N_\mathrm{T}$:
\begin{equation}
\int \d x \vecvar{\Phi}^{\dagger}(x,t) \vecvar{\Phi}(x,t) =N_\mathrm{T}.
\end{equation}
The spinor condensate wavefunction obeys a set of coupled evolution equations, namely, the multi-component GP equation:
\begin{eqnarray}
\label{spinorGP}
\im\hbar\p_t \Phi_1\eq -\frac{\hbar^2}{2m}\p^2_x \Phi_1+(\bar{c}_0+\bar{c}_2)\left(|\Phi_1|^2+|\Phi_{0}|^2\right)\Phi_1\nonumber \\
&&\hspace{-3mm}+(\bar{c}_0-\bar{c}_2)|\Phi_{-1}|^2\Phi_1+c_2\Phi^*_{-1}\Phi^2_0,\nonumber \\
\im\hbar\p_t \Phi_0\eq -\frac{\hbar^2}{2m}\p^2_x\Phi_0+(\bar{c}_0+\bar{c}_2)\left(|\Phi_1|^2+|\Phi_{-1}|^2\right)\Phi_0\nonumber \\
&&\hspace{-3mm}+\bar{c}_0|\Phi_0|^2\Phi_0+2\bar{c}_2\Phi^*_0\Phi_1\Phi_{-1},
\nonumber\\
\im\hbar\p_t\Phi_{-1}\eq -\frac{\hbar^2}{2m}\p^2_x\Phi_{-1}+(\bar{c}_0+\bar{c}_2)\left(|\Phi_{-1}|^2+|\Phi_{0}|^2\right)\Phi_{-1}\nonumber \\
&&\hspace{-3mm}+(\bar{c}_0-\bar{c}_2)|\Phi_{1}|^2\Phi_{-1}+\bar{c}_2\Phi^*_{1}\Phi^2_0,
\end{eqnarray}
where $\bar{c}_0=(\bar{g}_0+2\bar{g}_2)/3$ and $\bar{c}_2=(\bar{g}_2-\bar{g}_0)/3$ denote effective 1D coupling constants for the mean-field and the spin-exchange interaction, respectively. Here, the effective 1D coupling constants $\bar{g}_f$ are given by \cite{Olshanii}
\begin{equation}
\bar{g}_f=\frac{4\hbar^2a_f}{ma_\perp^2}\frac{1}{1-Ca_f/a_\perp},
\end{equation}
where $a_f$ are the $s$-wave scattering lengths in the total hyperfine spin $f$ channel, $a_\perp$ is the size of the transverse ground state, $m$ is the atomic mass, and $C=-\zeta(1/2)\simeq 1.46$. Note that one may change the values of $\bar{c}_0$ and $\bar{c}_2$ by tuning $a_\perp$. 
 
Equation (\ref{spinorGP}) is derived as follows. The interaction between two atoms in the $F=1$ hyperfine spin state has a form, \cite{Ho,OM}
\begin{equation}
\hat{V}(x_1-x_2)=\delta(x_1-x_2)\left(\bar{c}_0+\bar{c}_2\hat{\vecvar{F}}_1\cdot\hat{\vecvar{F}}_2\right),
\end{equation}
where $\hat{\vecvar{F}}_i$ is the spin operator. The Gross-Pitaevskii energy functional is thus given by
\begin{equation}
E_\mathrm{GP}[\vecvar{\Phi}] =  \int \d x \bigg( \frac{\hbar^2}{2m}\p_x\Phi_{\alpha}^{\ast}\p_x\Phi_\alpha+\frac{\bar{c}_0}{2}\Phi_{\alpha}^{\ast}\Phi_{\alpha'}^{\ast}\Phi_{\alpha'}\Phi_{\alpha}+\frac{\bar{c}_2}{2}\Phi_{\alpha}^{\ast}\Phi_{\alpha'}^{\ast}\textbf{f}_{\alpha\beta}^{\hspace{0.5mm} T}\cdot\textbf{f}_{\alpha'\beta'}\Phi_{\beta'}\Phi_{\beta} \bigg),
\end{equation}
where repeated subscripts ($\alpha, \beta, \alpha', \beta'=1,0,-1$) should be summed up and $\textbf{f}=(\textrm{f}^x, \textrm{f}^y, \textrm{f}^z)^T$ with $\textrm{f}^i (i=x,y,z)$ being $3\times3$ spin-1 matrices.
Then, the variational principle: $\im \hbar \p_t \Phi_\alpha (x,t) =\delta E_\mathrm{GP}[\vecvar{\Phi}]/\delta \Phi^\ast_\alpha (x,t)$, for $\alpha=1,0,-1$, yields eq. (\ref{spinorGP}). 

%Here, we use the following representation of spin-1 matrices $\textbf{f}=(\textrm{f}^x, \textrm{f}^y, \textrm{f}^z)^T$,
%\begin{equation}
%\label{su2}
%\mathrm{f}^x=\frac{1}{\sqrt{2}}\left(\begin{array}{ccc}
%0&1&0 \\
%1&0&1 \\
%0&1&0 \end{array}\right),
%\,
%\mathrm{f}^y=\frac{\i}{\sqrt{2}}\left(\begin{array}{ccc}
%0&-1& 0 \\
%1& 0&-1 \\
%0& 1& 0 \end{array}\right),
%\,
%\mathrm{f}^z=\left(\begin{array}{ccc}
%1&0& 0 \\
%0&0& 0 \\
%0&0&-1 \end{array}\right),
%\end{equation}

An important fact is that eq. (\ref{spinorGP}) possesses a completely integrable point when $\bar{c}_0=\bar{c}_2 \equiv -c <0$, equivalently, $2\bar{g}_0 =-\bar{g}_2>0$. \cite{IMW1, IMW2} This condition is realized when
\begin{equation}
a_{\perp}=3C\frac{a_0a_2}{2a_0+a_2},
\end{equation}
assuming that $a_0a_2(a_2-a_0)>0$ holds. The situation corresponds to attractive ($\bar{c}_0<0$) and ferromagnetic ($\bar{c}_2<0$) interaction. When we change the wavefunction by $\vecvar{\Phi}=(\phi_1,\sqrt{2}\phi_0,\phi_{-1})^T$ and measure time and length in units of $\bar{t}=\hbar a_\perp/c$ and $\bar{x}=\hbar\sqrt{a_\perp/2mc}$, respectively, we can rewrite eq. (\ref{spinorGP}) with $\bar{c}_0=\bar{c}_2 \equiv -c <0$ into the dimensionless form,
\begin{eqnarray}
\label{stdeq}
\im\partial_t \phi_1 \eq -\partial^2_{x}\phi_1
-2(|\phi_1|^2+2|\phi_0|^2)\phi_1-2\phi^*_{-1}\phi^2_0,\nonumber\\
\im\partial_t \phi_0 \eq -\partial^2_{x}\phi_0
-2(|\phi_{-1}|^2+|\phi_0|^2+|\phi_1|^2)\phi_0-2\phi^*_0\phi_1\phi_{-1},
\nonumber\\
\im\partial_t \phi_{-1} \eq -\partial^2_{x}\phi_{-1}
-2(|\phi_{-1}|^2+2|\phi_0|^2)\phi_{-1}-2\phi^*_{1}\phi^2_0.
\end{eqnarray}
Then, eq. (\ref{stdeq}) is found to be equivalent to a $2\times 2$ matrix version of the NLS equation with a self-focusing nonlinearity:
\begin{equation}
\label{NLS}
\im \partial_t Q+ \partial_x^2 Q +2QQ^{\dagger}Q=O,
\end{equation}
with an identification,
\begin{equation}
\label{reduction}
Q=\left(\hspace*{-1mm}\begin{array}{cc}
\phi_1 \hspace*{-1mm}&\hspace*{-1mm} \phi_0 \\
\phi_0 \hspace*{-1mm}&\hspace*{-1mm} \phi_{-1}
\end{array}\hspace*{-1mm}
\right).
\end{equation}
The matrix NLS equation (\ref{NLS}) is integrable in the sense that the initial value problem can be solved via the inverse scattering method \cite{Tsuchida, ISM_NVBC}. The integrability of the reduced equations (\ref{stdeq}) is thus proved automatically. Thus, we have derived the integrable spinor model. Another integrable point of eq. (\ref{spinorGP}) is $\bar{c}_0=\bar{c}_2\equiv c>0$, i.e., the matrix NLS equation with a self-defocusing nonlineality \cite{UIW}. Special solutions for generic coupling constants $\bar{c}_0$, $\bar{c}_2$ are given in ref. \citen{WT}. 

\section{Bright Solitons with a Finite Background}
\label{sec:bright soliton under NVBC}

We consider bright soliton solutions of the integrable model (\ref{stdeq}) under NVBC, whereas those under VBC are studied in refs. \citen{IMW1} and \citen{IMW2}. We summarize briefly the results of the inverse scattering method for eq. (\ref{stdeq}) with NVBC \cite{ISM_NVBC}.  

We define the nonvanishing boundary conditions as
\begin{align}
\label{constBC}
Q(x,t)\to Q_\pm, \hspace{0.5cm}x\to\pm\infty,&\nonumber \\
Q_\pm^\dagger Q_\pm= Q_\pm Q_\pm^\dagger=\lambda_0^2I,&
\end{align}
where $\lambda_0$ is a positive real constant and $I$ denotes a $2\times 2$ unit matrix. Note that vanishing boundary conditions are recovered as $\lambda_0\to 0$. The analysis of the ISM under NVBC yields the standard form of the multiple soliton solutions of the $2\times2$ matrix NLS equation with a self-focusing nonlineality (\ref{NLS}) as
\begin{equation}
Q(x,t)=\lambda_0\e^{\smalli\phi}
\left\{I+2\im(\underbrace{I\cdots I}_N)S^{-1}\left(\hspace*{-1mm}
\begin{array}{c}
\Pi_1\e^{\chi_1}\\
\vdots\\
\Pi_{N}\e^{\chi_{N}}
\end{array}
\hspace*{-1mm}\right)\right\}.
\label{Nsoliton}
\end{equation}
Here, a $2\times 2$ complex matrix $\Pi_j$ is called the polarization matrix. $S$ is a $2N\times 2N$ matrix defined by
\begin{align}
\label{Smat}
S_{ij}=\frac{\lambda_0}{\zeta_j+\lambda_j}\delta_{ij}I
+\frac{\lambda_0^2}{\im(\zeta_i+\zeta_j)}
\left(\frac{1}{\zeta_i+\lambda_i}+\frac{1}{\zeta_j+\lambda_j}\right)\Pi_i\e^{\chi_i},& \nonumber \\
1\le i,j\le N,&
\end{align}
where $\lambda_j$ is a complex discrete eigenvalue for the bound state and $\zeta_j=(\lambda_j^2+\lambda_0^2)^{1/2}$ with $\mathrm{Im} \ \zeta_j>0$ for $j=1,\dots,N$. It is required for the ISM under NVBC that a two-sheet Riemann surface is introduced appropriately, due to a double-valued function $\zeta_j$.
The phase of the carrier wave, $\phi(x,t)$, is given by
\begin{equation}
\phi(x,t)=kx-(k^2-2\lambda_0^2)t+\delta,
\end{equation}
and the coordinate function is given by
\begin{equation}
\chi_j\equiv\chi_j(x,t)=2\im\zeta_j(x-2(\lambda_j+k)t).
\end{equation}

The above solution is the $M (=N/2)$-soliton solution. The ISM under NVBC for the self-focusing case results in pairs of discrete eigenvalues corresponding to each Riemann sheet. 
The constraint should be imposed on $\lambda_j$ and $\zeta_j$ ($j=1,\cdots,N$) 
such that $\lambda_{2l-1}=\lambda_{2l}^*$ and $\zeta_{2l-1}=-\zeta_{2l}^*$ for $l=1,\cdots,N/2$. 
At the same time, $\Pi_j$ must satisfy that $\Pi_{2l-1}=\Pi_{2l}^\dagger$.

For our reduction to the integrable model for $F=1$ spinor BEC, we must make the potential $Q$ symmetric, noting eq. (\ref{reduction}). The symmetry of $Q$ is naturally reflected in $\Pi_j$. When we take every $\Pi_j$ to be symmetric in eq. (\ref{Nsoliton}), soliton solutions of the integrable model (\ref{stdeq}) under NVBC are obtained.

The form (\ref{Nsoliton}) is the standard form of soliton solutions in the sense that the boundary value at $x\to \infty \Leftrightarrow \e^{\chi_j}\to 0$ is supposed to be fixed as
\begin{equation}
\label{limit}
Q(x,t)\e^{-\smalli\phi(x,t)}\to \lambda_0 I, \hspace{0.5cm} x\to\infty.
\end{equation}
The spinor model, however, allows the $SU(2)$ transformation of the solutions, if they are kept symmetric. To be concrete, let $\mathcal{U}$ be a $2\times 2$ unitary matrix. When $Q$ is a solution of eq. (\ref{NLS}) with eq. (\ref{reduction}), then
\begin{equation}
\label{rotation}
Q'=\mathcal{U}Q\mathcal{U}^T,
\end{equation}
is also a solution. Assuming that $Q$ is the standard form (\ref{Nsoliton}), the limit $x\to \infty$ of $Q'$ becomes
\begin{equation}
Q'\e^{-\smalli\phi(x,t)}\to \lambda_0 \ \mathcal{U}\mathcal{U}^T \equiv \lambda_0 Q'_+, \hspace{0.5cm} x\to\infty.
\end{equation}
$Q'_+=\mathcal{U}\mathcal{U}^T$ is the so-called Cholesky decomposition. The arbitrary boundary conditions $Q'_+$ other than eq. (\ref{limit}) are thus realized via the $SU(2)$ transformation. On the other hand, the behavior in the limit $x\to -\infty$ varies depending on whether $\det \Pi=0$ or not, which will be discussed later for the one-soliton case. 

There is another important concept about an integrable model. Due to the integrability, the model has the infinite conservation laws, which restrict the dynamics of the system in an essential way. Several conserved quantities, related to the physical quantities of the system, are listed below:
\begin{eqnarray}
\textbf{Total number} \hspace{-3mm}&:&\hspace{-3mm} \bar{N}_{\mathrm{T}}=\int \d x \ \bar{n}(x,t), \nonumber\\
&&\hspace{-20mm} \bar{n}(x,t)=\tr(Q^{\dagger}Q)-\tr(Q_{\pm}^{\dagger}Q_{\pm}). \label{tnumber}\\
\textbf{Total spin} \hspace{-3mm}&:&\hspace{-3mm} \vecvar{F}_{\mathrm{T}}=\int \d x \ \vecvar{f}(x,t), \nonumber\\
&&\hspace{-20mm} \vecvar{f}(x,t)= \tr(Q^{\dagger}\vecvar{\sigma}Q). \label{tspin}\\
\textbf{Total momentum} \hspace{-3mm}&:&\hspace{-3mm} \bar{P}_{\mathrm{T}}=\int \d x \ \bar{p}(x,t), \nonumber\\
&&\hspace{-20mm} \bar{p}(x,t)=-\im\hbar[\tr(Q^{\dagger}Q_x)-\tr(Q_{\pm}^{\dagger}Q_{\pm,x})]. \label{tmomuntum}\\
\textbf{Total energy} \hspace{-3mm}&:&\hspace{-3mm} \bar{E}_{\mathrm{T}}=\int \d x \ \bar{e}(x,t), \nonumber\\
&&\hspace{-20mm} \bar{e}(x,t)=c[\tr(Q^{\dagger}_xQ_x-Q^{\dagger}QQ^{\dagger}Q)\nonumber\\
&&\hspace{-20mm} \ \ \ \ \ \ \ \ \ \ \ \ \ \ -\tr(Q^{\dagger}_{\pm,x}Q_{\pm,x}-Q^{\dagger}_{\pm}Q_{\pm}Q^{\dagger}_{\pm}Q_{\pm})]. \label{tenergy}
\end{eqnarray}
Here, $\vecvar{\sigma}=( \sigma_x, \sigma_y, \sigma_z )^T$ are the Pauli matrices. 
%defined by
%\begin{equation}
%\sigma_x= \left(\hspace*{-1mm}\begin{array}{cc}
%0 \hspace*{-1mm}&\hspace*{-1mm}1\\
%1\hspace*{-1mm}&\hspace*{-1mm}0
%\end{array}\hspace*{-1mm}\right), \hspace{0.5cm}
%\sigma_y= \left(\hspace*{-1mm}\begin{array}{cc}
%0 \hspace*{-1mm}&\hspace*{-1mm}-\im\\
%\im\hspace*{-1mm}&\hspace*{-1mm}0
%\end{array}\hspace*{-1mm}\right), \hspace{0.5cm}
%\sigma_z= \left(\hspace*{-1mm}\begin{array}{cc}
%1 \hspace*{-1mm}&\hspace*{-1mm}0\\
%0\hspace*{-1mm}&\hspace*{-1mm}-1
%\end{array}\hspace*{-1mm}\right).
%\end{equation}
To avoid the divergence of integrals, we should subtract the contribution of the background from the physical quantities, except the total spin, to which the background does not contribute explicitly. These subtractions are emphasized by the bars on the conserved densities and quantities. The local spin density $\vecvar{f}=(f_x, f_y, f_z)^T$ is 
covariant under the $SU(2)$ transformation (\ref{rotation}), whereas the other densities such as $\bar{n}(x,t)$, $\bar{p}(x,t)$ and $\bar{e}(x,t)$ 
are invariant. The total macroscopic spin is directed to face to an arbitrary direction by the global spin rotation. The $SU(2)$ symmetry of the system causes the energy degenerated states of solitons for this spin rotation. 

\section{One-Soliton Solutions}
\label{sec:one-soliton}

In this section, one-soliton solutions of the integrable spinor model (\ref{stdeq}) are investigated in detail. We can derive the explicit form of one-soliton solutions by setting $N=2$ ($M=1$) in the formula (\ref{Nsoliton}). The calculation is complicated but straightforward. The result is as follows:
\begin{equation}
\label{NVBC1} 
Q=\lambda_0 \e^{\smalli \phi(x,t)}\left( I +2\im \, \frac{T}{\det S} \right),
\end{equation}
where $\det S$ is given by
\begin{eqnarray}
\label{NVBC2} 
\det S \eq \k_1^2\k_2^2-\e^{\chi_1}\nu_1\k_1\k_2^2\,\tr \Pi -\e^{x_2}\nu_2\k_1^2\k_2 \,\tr \Pi^{\dagger} \nonumber \\
&&\hspace{-3mm} +\e^{\chi_1+\chi_2}\k_1\k_2\left\{ \varpi+\nu_1\nu_2(|\tr \Pi|^2-1) \right\} \nonumber \\
&&\hspace{-3mm} +\e^{2\chi_1}\nu_1^2\k_2^2\,\det \Pi +\e^{2\chi_2}\nu_2^2\k_1^2 \,\det \Pi^{\dagger} \nonumber \\
&&\hspace{-3mm} -\e^{2\chi_1+\chi_2}\nu_1\k_2 \varpi\, \tr \Pi^{\dagger} \,\det \Pi - \e^{\chi_1+2\chi_2}\nu_2\k_1 \varpi\,\tr \Pi \,\det \Pi^{\dagger}\nonumber \\
&&\hspace{-3mm} +\e^{2\chi_1+2\chi_2}\varpi^2|\det \Pi|^2,
\end{eqnarray}
and $T$ is a $2\times 2$ matrix such that
\begin{eqnarray}
\label{NVBC3} 
T\eq \e^{\chi_1}\k_1\k_2^2\, \Pi+\e^{\chi_2}\k_1^2\k_2\, \Pi^{\dagger} \nonumber \\
&&\hspace{-3mm} -\e^{\chi_1+\chi_2}\k_1\k_2 \left\{ \varsigma_1\,\tr \Pi \cdot \Pi^{\dagger} + \varsigma_2\,\tr \Pi^{\dagger} \cdot \Pi +\mu (|\tr \Pi|^2-1)I \right\} \nonumber \\
&&\hspace{-3mm} -\e^{2\chi_1}\mu_1\k_2^2\,\det \Pi \cdot I-\e^{2\chi_2}\mu_2\k_1^2\,\det \Pi^{\dagger} \cdot I \nonumber \\
&&\hspace{-3mm} +\e^{2\chi_1+\chi_2}\k_2 \,\det \Pi \left\{ \varsigma_1^2 \,\Pi^{\dagger} +\varpi\,\tr \Pi^{\dagger} \cdot I \right\} \nonumber \\
&&\hspace{-3mm} +\e^{\chi_1+2\chi_2}\k_1 \,\det \Pi^{\dagger}\left\{ \varsigma_2^2 \,\Pi +\varpi\,\tr \Pi \cdot I \right\} \nonumber \\
&&\hspace{-3mm} -\e^{2\chi_1+2\chi_2}\varpi\left( \varsigma_1+\varsigma_2 \right)|\det \Pi|^2 \cdot I.
\end{eqnarray}
We explain physical meanings of notations. The phase of the carrier wave is given by
\begin{equation}
\phi(x,t)=kx-(k^2-2\lambda_0^2)t+\delta.
\end{equation}
Let $\lambda_j$ and $\zeta_j=(\lambda_j^2+\lambda_0^2)^{1/2}$ for $j=1,2$ be complex constants satisfying $\lambda_1=\lambda_2^*$ and $\zeta_1=-\zeta_2^*$. Without loss of generality, we assume that $(\lambda_1,\zeta_1) \ ((\lambda_2,\zeta_2))$ belongs to the upper (lower) Riemann sheet, which is characterized such that $\mathrm{Im}\ \zeta\ \mathrm{Im}\ \lambda>0 \ (\mathrm{Im}\ \zeta\ \mathrm{Im}\ \lambda<0)$. $\chi_j(x,t)$ is expressed in terms of them as
\begin{eqnarray}
\chi_1\eequiv\chi_1(x,t)=2\im\zeta_1(x-2(\lambda_1+k)t),\\
\chi_2\eequiv\chi_2(x,t)=2\im\zeta_2(x-2(\lambda_2+k)t).
\end{eqnarray}
Note that $\chi_1=\chi_2^*\equiv \chi$ holds. $\mathrm{Re}\,\chi$ thus denotes the coordinate of the envelope soliton, whereas $\mathrm{Im}\,\chi$ implies the self-modulation phase. The polarization matrix $\Pi$ is a $2\times2$ symmetric matrix. Here, the normalization in a sense of the square norm is imposed on $\Pi$ as a matter of convenience:
\begin{equation}
\Pi\equiv \left(\hspace*{-1mm}\begin{array}{cc}
\beta \hspace*{-1mm}&\hspace*{-1mm}\alpha\\
\alpha\hspace*{-1mm}&\hspace*{-1mm}\gamma
\end{array}\hspace*{-1mm}\right), \hspace{0.5cm} 2|\alpha|^2+|\beta|^2+|\gamma|^2=1.
\end{equation}
The other parameters are expedient functions of $\lambda_0, \lambda_1$ and $\lambda_2$: 
\begin{equation}
\k_j = \frac{\lambda_0}{\zeta_j+\lambda_j}, \ \ \mu=\im\lambda_0\frac{\k_1+\k_2}{\zeta_1+\zeta_2}, \ \ \nu_j=\frac{\im\lambda_0\k_j}{\zeta_j}, \ \ \varpi=\nu_1\nu_2-\mu^2, \ \ \varsigma_j=\nu_j-\mu,
\end{equation}
for $j=1,2$. We list the meaning of each parameter as follows:
\begin{eqnarray*}
k\hspace{-2mm}&:&\hspace{-2mm} \textrm{wave number of soliton's carrier wave}.\\
\lambda_0\hspace{-2mm}&:&\hspace{-2mm} \textrm{amplitude of soliton's carrier wave}.\\
\phi(x,t)\hspace{-2mm}&:&\hspace{-2mm} \textrm{phase of soliton's carrier wave}.\\
\mathrm{Re} \ \chi(x,t) \hspace{-2mm}&:&\hspace{-2mm} \textrm{coordinate of soliton's envelope}.\\
\mathrm{Im} \ \chi(x,t) \hspace{-2mm}&:&\hspace{-2mm} \textrm{self-modulation phase of soliton}.\\
\Pi \hspace{-2mm}&:&\hspace{-2mm} \textrm{symmetric polarization matrix of soliton}.
\end{eqnarray*}

Equations (\ref{NVBC1})-(\ref{NVBC3}) are new soliton solutions which had never been written down explicitly in the literatures. If we take the vanishing limit $\lambda_0 \to 0$, $\zeta_1$ and $\zeta_2$ converge at $\lambda_1$ and $-\lambda_2$, respectively. Then,
\begin{equation}
\label{VBC1}
Q \to \left(\im\frac{\lambda_1}{|\lambda_1|}\e^{\smalli\delta}\right)\cdot 2k_R\ \frac{\Pi\e^{-(\chi_R+\rho/2)}+(\sigma^{y}\Pi^{\dagger}\sigma^y)\e^{\chi_R+\rho/2}\det\Pi}{\e^{-(2\chi_R+\rho)}+1+\e^{2\chi_R+\rho}|\det\Pi|^2}\ \e^{\smalli\chi_I},
\end{equation}
with notations
\begin{eqnarray}
\e^{\rho/2}\hspace{-2mm}&\equiv&\hspace{-2mm} \frac{1}{2k_R}, \label{VBC11}\\
\chi_R\hspace{-2mm}&\equiv&\hspace{-2mm} \chi_R(x,t)=k_R(x-2k_It)-\epsilon, \\
\chi_I\hspace{-2mm}&\equiv&\hspace{-2mm} \chi_I(x,t)=k_Ix+(k_R^2-k_I^2)t\label{VBC12},
\end{eqnarray}
each of which holds the following correspondence respectively:
\begin{eqnarray}
k_R\eq -2\textrm{Im} \lambda_1,\\
k_I\eq 2\textrm{Re} \lambda_1+k,\\
\epsilon\eq -\ln (4|\lambda_1|)\label{initial}.
\end{eqnarray}
Equations (\ref{VBC1})-(\ref{VBC12}) are the same forms as those in ref. \citen{IMW2}, except a phase factor. This consequence is natural but non-trivial, because the formula of solitons under VBC \cite{Tsuchida} is quite different from that under NVBC (\ref{Nsoliton}), in particular, in the form of the matrix $S$. Actually, the initial displacement $\epsilon$ can be arbitrarily changed, regardless of eq. (\ref{initial}), by the parallel shift of the position $x$. We have shown that the soliton solutions (\ref{NVBC1})-(\ref{NVBC3}) can be regarded as a general form of bright soliton solutions, including the case of VBC. 

\subsection{Classification by the boundary conditions}

\begin{figure}[htbp]
\unitlength=1mm
\begin{center}
\begin{picture}(150,78)(0,-3)
\put(19,-6){(a)}
\put(71,-6){(b)}
\put(125,-6){(c)}
\begin{minipage}[b]{50mm}
\includegraphics[scale=.38]{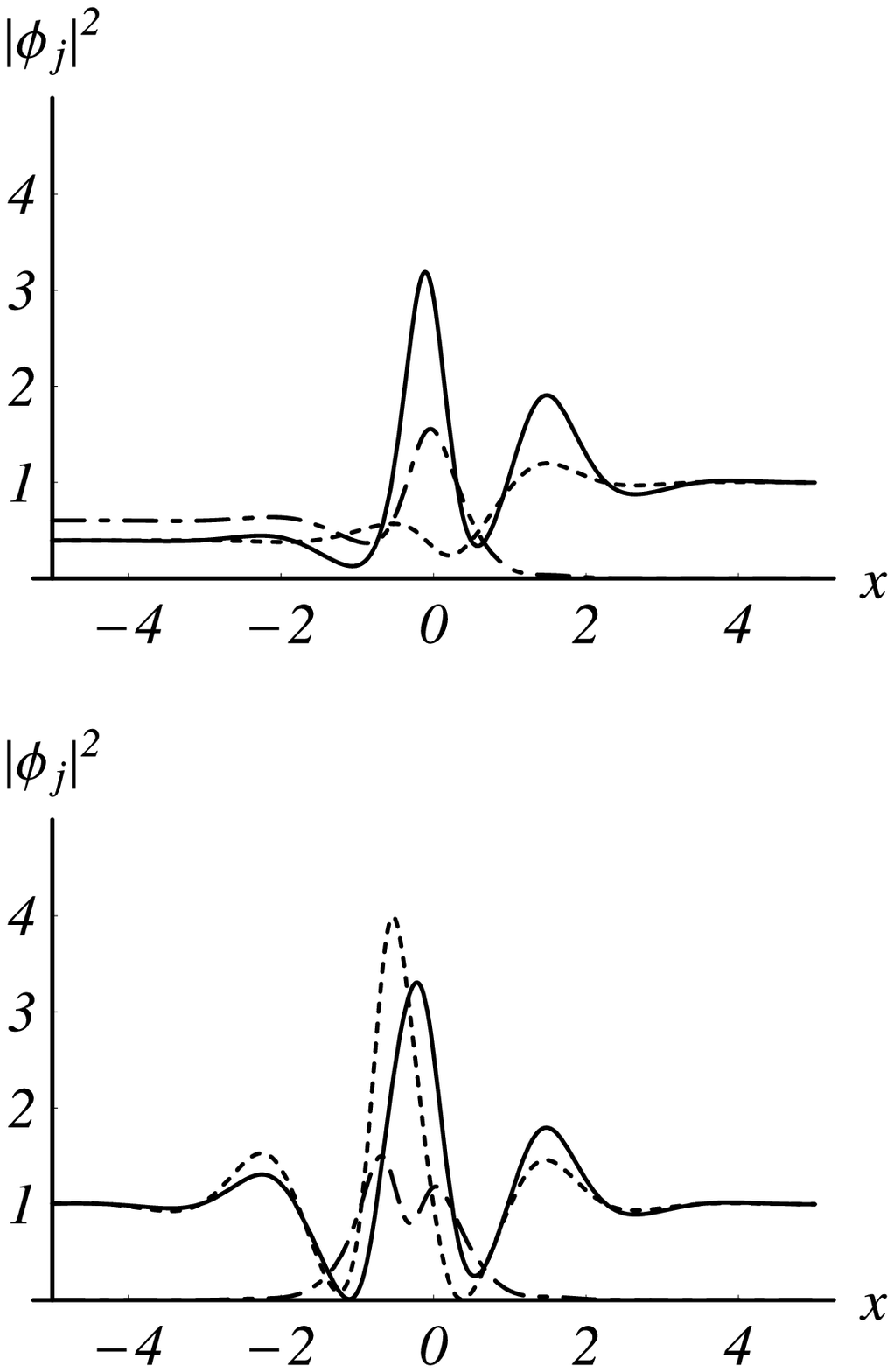}
\end{minipage}
\begin{minipage}[b]{50mm}
\includegraphics[scale=.38]{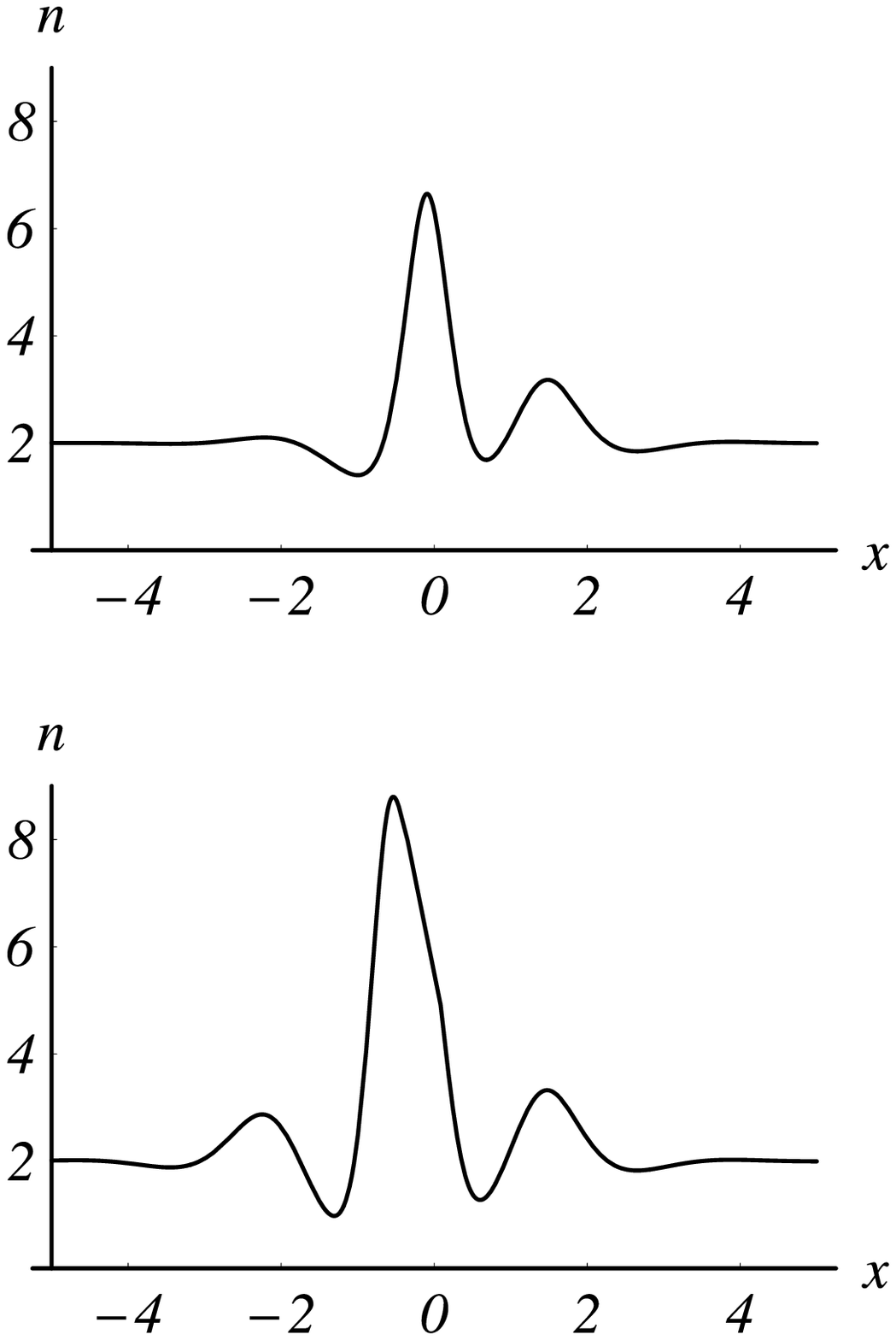}
\end{minipage}
\begin{minipage}[b]{50mm}
\includegraphics[scale=.38]{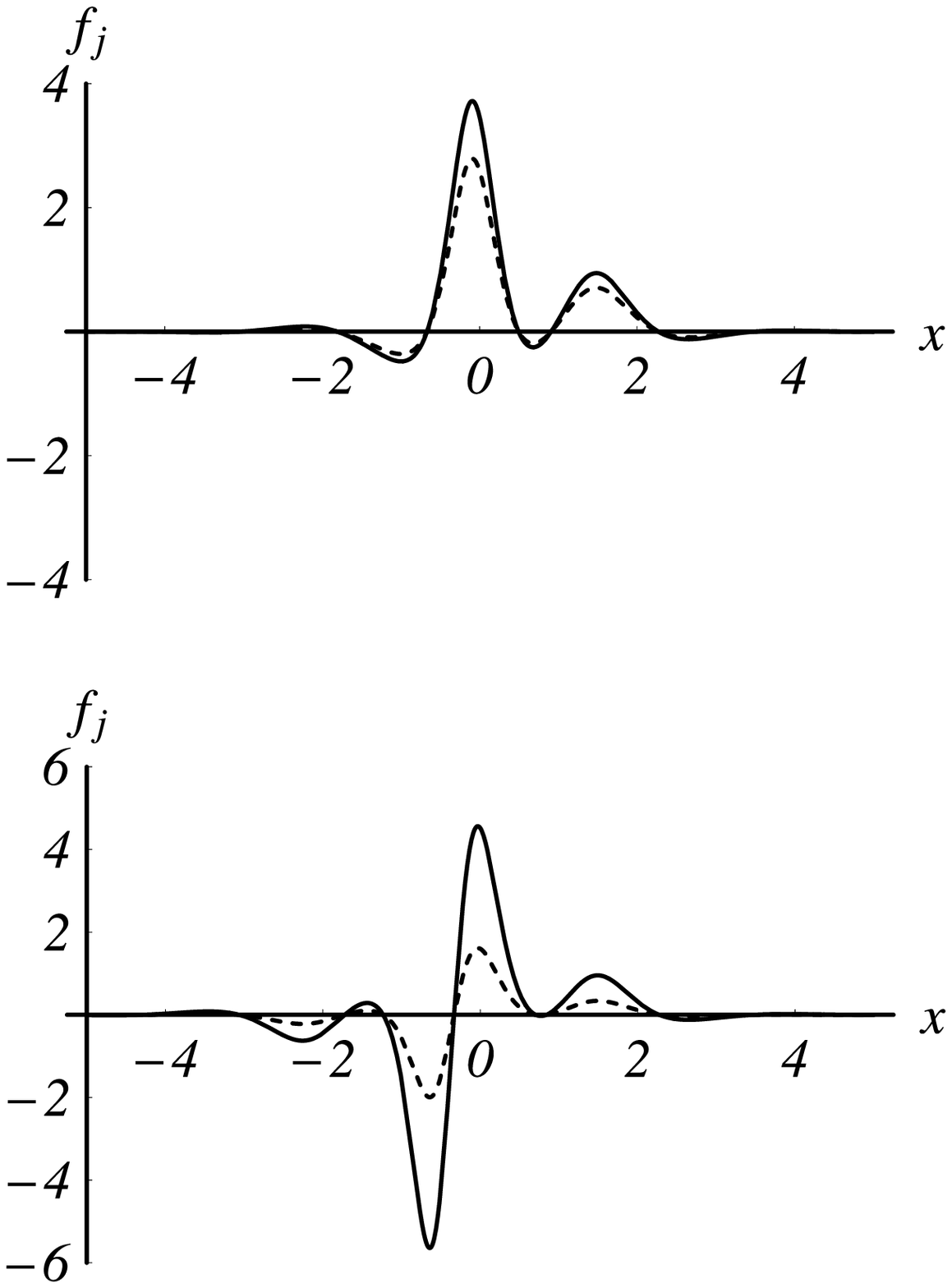}
\end{minipage}
\end{picture}
\end{center}
\caption{Snapshots of one-soliton density profiles. The upper row is plotted for $\det \Pi=0$ at the moment $t=0$, with $k=0$, $\lambda_0=1$, $\lambda_1=1+\im$, $\xi_1=1.27+0.79\im$ ($\chi(x,t)=-(1.57-2.54\im)x-(8.23-1.94\im)t$) and $\Pi=\left(\hspace*{-1mm}\begin{array}{cc}
4/5 \hspace*{-1mm}&\hspace*{-1mm}2/5\\
2/5\hspace*{-1mm}&\hspace*{-1mm}1/5
\end{array}\hspace*{-1mm}\right)$. The lower row is plotted for $\det \Pi\neq 0$ at the moment $t=0$, with the same parameters except for $\Pi=\left(\hspace*{-1mm}\begin{array}{cc}
1/\sqrt{2} \hspace*{-1mm}&\hspace*{-1mm}2/5\\
2/5\hspace*{-1mm}&\hspace*{-1mm}3/(5\sqrt{2})
\end{array}\hspace*{-1mm}\right)$. The left panel (a) depicts the local density for each component, $|\phi_1|^2$ (solid line), $|\phi_0|^2$ (chain line) and $|\phi_{-1}|^2$ (dotted line). The center panel (b) depicts the local number density $n$, where the contribution of the background is included. The right panel (c) depicts the local spin densities, $f_x$ (solid line) and $f_z$ (dotted line). $f_y$ vanishes identically due to a choice of a real matrix $\Pi$. 
}
\label{fig:1sol}
\end{figure}

We shall show that there are two kinds of one-soliton solutions depending on the boundary conditions, $\det \Pi=0$ or $\det \Pi\neq 0$. The similar classification about the boundary conditions also exists for dark solitons \cite{UIW}. The examples of snapshots of one-soliton density profiles are shown in Fig. \ref{fig:1sol}. The upper row is for $\det \Pi=0$, and the lower row is for $\det \Pi\neq 0$. The shape of envelope solitons looks a locally-oscilating wave rather than, literally, a solitary wave, because of the self-modulation due to the complex velocity.

For $\det \Pi=0$, the boundary conditions of the standard form (\ref{NVBC1})-(\ref{NVBC3}) are
\begin{align}
Q\e^{-\smalli\phi}&\to \lambda_0 I, & x&\to\infty,\nonumber\\
Q\e^{-\smalli\phi}&\to \lambda_0 \left[ I-2\im\, \frac{\varsigma_1\,\tr \Pi\cdot \Pi^{\dagger}+\varsigma_2\, \tr \Pi^{\dagger}\cdot \Pi+\mu(|\tr \Pi|^2-1)I}{\varpi+\nu_1\nu_2(|\tr \Pi|^2-1)} \right], & x&\to -\infty.
\end{align}
The left and right boundary values differ in not only the global phase but also the population of each component, in general. That is, those are the $SU(2)$ rotated boundary conditions. In the upper row of Fig. \ref{fig:1sol}, we see that the envelope soliton of each component forms the domain-wall (DW) shape, although it does not manifest in the total number density.

On the other hand, for $\det \Pi\neq 0$, the boundary conditions are  
\begin{align}
Q\e^{-\smalli\phi}&\to \lambda_0 I, & x&\to\infty,\nonumber\\ 
Q\e^{-\smalli\phi}&\to \lambda_0 \left(1-2\im\,\frac{\varsigma_1+\varsigma_2}{\varpi}\right)I, & x&\to -\infty.
\end{align}
In contrast to the case that $\det \Pi=0$, both boundary values are diagonal matrices, and only the phase-shift (PS) occurs. That is, those are the $U(1)$ rotated boundary conditions.

For the above reasons, we call one-soliton solutions the DW-type for $\det \Pi =0$, and the PS-type for $\det \Pi \neq 0$. Remark the following; the spin density profile of DW-type suggests that the total spin is nonzero, whereas that of PS-type is dipole-shape, implying that the total spin amounts to zero. See the right panel of Fig. \ref{fig:1sol}. This observation will be solidified in $\S$ \ref{subsec:spin}.

\subsection{Case of purely imaginary discrete eigenvalues}

The ISM performed on Riemann sheets involves a double-valued function of the spectral parameter, and it usually renders a very complicated representation of $N$-soliton solutions even for $N=1$, as is seen from eqs. (\ref{NVBC1})-(\ref{NVBC3}). To simplify an explicit representation, it is convenient to assume that $\lambda_j$ and $\zeta_j$ are purely imaginary \cite{KI78}. The similar approach is employed for the analysis of $N$-soliton solutions of the derivative NLS equation under NVBC \cite{dNLS}.

If we take a pair of discrete eigenvalues as
\begin{equation}
(\lambda_1, \zeta_1) \equiv (\im\lambda_0\lambda, \im\lambda_0\zeta), \hspace{0.5cm} (\lambda_2, \zeta_2) \equiv (-\im\lambda_0\lambda, \im\lambda_0\zeta),
\end{equation}
where $\lambda$ and $\zeta$ are positive real numbers such that
\begin{equation}
\lambda>1, \hspace{0.5cm} \zeta =\sqrt{\lambda^2-1},
\end{equation}
we obtain a relatively simple form of one-soliton solutions as
%\begin{allowdisplaybreaks}
\begin{eqnarray}
Q\eq\lambda_0 \e^{\smalli \phi(x,t)}\left( I +2\im \, \frac{T}{\det S} \right),\label{PNVBC1}\\
\det S\eq1-\frac{\e^{\chi_P}}{\zeta}(\tr \Pi(t)+\tr \Pi^{\dagger} (t))+\e^{2\chi_P}\left( 1+\frac{1}{\zeta^2}|\tr\Pi(t)|^2 \right)\nonumber \\
&&\hspace{-3mm}+\frac{\e^{2\chi_P}}{\zeta^2}(\det \Pi(t)+\det \Pi^{\dagger}(t))-\frac{\lambda^2}{\zeta^3}\e^{3\chi_P}\left(\tr \Pi^{\dagger}(t)\det \Pi(t)+\tr\Pi (t)\det \Pi^{\dagger}(t)\right)\nonumber\\
&&\hspace{-3mm}+\frac{\lambda^4}{\zeta^4}\e^{4\chi_P}|\det \Pi(t)|^2,\label{PNVBC2}\\
2\im \ T\eq\Bigg\{ 2(|\tr\Pi(t)|^2-1)\e^{2\chi_P}+2\e^{2\chi_P}\left[\left(1+\frac{\lambda}{\zeta}\right)\det \Pi(t)+\left( 1-\frac{\lambda}{\zeta} \right)\det \Pi^{\dagger}(t)\right] \nonumber\\
&&\hspace{-3mm} \ \ \ \ \ \ \ \ \ \ -2\frac{\lambda^2}{\zeta}\e^{3\chi_P}\left[\left( 1+\frac{\lambda}{\zeta} \right) \tr \Pi^{\dagger}(t)\det \Pi (t)+\left(1-\frac{\lambda}{\zeta}\right)\tr \Pi (t)\det \Pi^{\dagger}(t)\right]\Bigg\}I\nonumber\\
&&\hspace{-3mm}+\left[-2(\zeta+\lambda)\e^{\chi_P}+2\frac{\lambda}{\zeta}\e^{2\chi_P}\tr \Pi^{\dagger}(t)+2\frac{\lambda^2}{\zeta}\left( 1-\frac{\lambda}{\zeta}\right)\e^{3\chi_P}\det \Pi^{\dagger}(t) \right]\Pi (t)\nonumber \\
&&\hspace{-3mm}+\left[ -2(\zeta-\lambda)\e^{\chi_P}-2\frac{\lambda}{\zeta}\e^{2\chi_P}\tr \Pi(t) +2\frac{\lambda^2}{\zeta}\left( 1+\frac{\lambda}{\zeta}\right)\e^{3\chi_P}\det \Pi(t) \right]\Pi^{\dagger}(t),\label{PNVBC3}
\end{eqnarray}
%\end{allowdisplaybreaks}
where the coordinate of the envelope soliton is given by
\begin{equation}
\chi_P(x,t)=-2\lambda_0 \zeta(x-2kt),\label{PNVBC11}
\end{equation}
and the time dependence of the phase modulation is embedded in the polarization matrix, namely,
\begin{equation}
\Pi(t)\equiv \Pi\e^{4\smalli \lambda_0^2\zeta\lambda t}.\label{PNVBC12}
\end{equation}
When we take the limit $\lambda_0\to 0$ with $\lambda_0\lambda$ and $\lambda_0\zeta$ kept finite in eqs. (\ref{PNVBC1})-(\ref{PNVBC3}), $Q$ converges to the form of eqs. (\ref{VBC1})-(\ref{VBC12}), accompanying the parameters $k_R=-2\lambda_0\lambda$, $k_I=k$ and $\epsilon=-\ln (4\lambda_0\lambda)$. Here, $k_R$ and $k_I$ are independent free parameters, apart from the trivial initial displacement $\epsilon$. In this sense, in spite of the reduction, we can still regard the form of one-soliton solutions (\ref{PNVBC1})-(\ref{PNVBC3}) as a general form of those under VBC.

We can also take another limit. That is, we consider the reduction to the single-component case. If we set 
\begin{equation}
k=0, \hspace{0.5cm} \Pi=\left(\hspace*{-1mm}\begin{array}{cc}
\e^{\smalli\theta} \hspace*{-1mm}&\hspace*{-1mm}0\\
0\hspace*{-1mm}&\hspace*{-1mm}0
\end{array}\hspace*{-1mm}\right),
\end{equation}
the (1,1)-component of $Q$ becomes
\begin{equation}
\label{KW_sol}
Q_{11} \cdot\e^{-\smalli(2\lambda_0^2 t +\delta)}= \lambda_0 - 2\lambda_0\zeta\, \frac{\zeta\cos (4\lambda_0^2\zeta\lambda t+\theta)+\im\lambda\sin (4\lambda_0^2\zeta\lambda t+\theta)}{\lambda \cosh (2\lambda_0 \zeta x+\psi)-\cos (4\lambda_0^2\zeta\lambda t+\theta)},
\end{equation}
where $\e^{\psi}=\zeta/\lambda$.
The form (\ref{KW_sol}) was given in ref. \citen{KI78}. We thus verify that our soliton solutions are also generalization of those for the single-component NLS equation under NVBC\cite{revise}. 

\subsection{Spin states}
\label{subsec:spin}

In this subsection, we discuss the spin states of one-soliton solutions with a finite background, by calculating the total spin. The conservation law guarantees that we obtain the total spin $\vecvar{F}_\mathrm{T}$ from integrating at arbitrary time. Therefore, we can select the time so that the calculation becomes easier. We concentrate on the case of purely imaginary discrete eigenvalues. As a result, we see that the DW-type is associated with the ferromagnetic state, whereas the PS-type is associated with the polar state. One-soliton solutions for purely imaginary discrete eigenvalues (\ref{PNVBC1})-(\ref{PNVBC3}) include those under VBC apart from an initial displacement, and therefore the classification about the spin states presented below is wider than that performed before \cite{IMW1, IMW2}.

\subsubsection{Ferromagnetic state}

For $\det\Pi=0$ (DW-type), we substitute eqs. (\ref{PNVBC1})-(\ref{PNVBC3}) into eq. (\ref{tspin}) and calculate the total spin. The time $t'$ such that $\tr \Pi(t')+\tr \Pi^{\dagger}(t')=0$ is suitable for the calculation. The result is as follows:
\begin{equation}
\vecvar{F}_{\mathrm{T}} = 4\lambda_0 \tau \ \frac{\lambda}{\zeta}\left(
\hspace*{-1mm}\begin{array}{c}
2\lambda\mathrm{Re}\{ \alpha^{\ast}(\beta+\gamma) \} \\
-2\zeta\mathrm{Im}\{ \alpha^{\ast}(\beta-\gamma) \}\\
\lambda(|\beta|^2-|\gamma|^2)\\
\end{array}\hspace*{-1mm}
\right),\hspace{0.5cm} \tau\equiv \left( 1+\frac{1}{\zeta^2}|\beta+\gamma|^2 \right)^{-1},
\end{equation}
with the modulus
\begin{equation}
|\vecvar{F}_\mathrm{T}|^2=(4\lambda\lambda_0)^2\tau.
\end{equation}
The total number of the particles transformed into a soliton, $\bar{N}_\mathrm{T}$, is calculated by eq. (\ref{tnumber}),
\begin{equation}
\bar{N}_\mathrm{T}=4\lambda_0\zeta,
\end{equation}
and the range of the value taken by $|\vecvar{F}_\mathrm{T}|^2$ is expressed in terms of $\bar{N}_\mathrm{T}$,
\begin{equation}
\bar{N}_{\mathrm{T}}^2 \le |\vecvar{F}_{\mathrm{T}}|^2 \le \bar{N}_{\mathrm{T}}^2+(4\lambda_0)^2.
\end{equation}
Remark that, in the vanishing limit $\lambda_0\to 0$, the modulus of the total spin is always equal to the total number of the particles, namely, $|\vecvar{F}_\mathrm{T}|\to N_\mathrm{T}$.

With nonzero total spin, the DW-type of solitons belongs to the ferromagnetic state. Since inter-atomic ferromagnetic interactions are supposed here, solitons tend to take the ferromagnetic state or DW-type. In various contexts of physics, the domain-walls are topological solitons related with the symmetry breaking. Here, resulting from the domain-walls, the magnetic entity emerges as the spontaneously broken symmetry. It is worthy to notice that the case $|\bar{N}_\mathrm{T}|<|\vecvar{F}_\mathrm{T}|$ may happen. The background is spinless, but its internal spin state appears to be affected on the ground that the ferromagnetic soliton runs over the background. Thereby, the background contributes to the total spin. 

\subsubsection{Polar state}

If $\det\Pi\neq 0$ (PS-type), the solitons show the other magnetic property. The time $t'$ such that $\det\Pi(t')=\det\Pi^{\dagger}(t')>0$ is suitable for the analysis. After lengthy calculations, the local spin density at such time is derived as
\begin{eqnarray}
\left(
\hspace*{-1mm}\begin{array}{c}
f_x \\
f_z\\
\end{array}\hspace*{-1mm}
\right)
\eq 8\lambda_0^2\e^{-3\upsilon}\Xi^{-2}(\chi_{P'})\left(
\hspace*{-1mm}\begin{array}{c}
2\alpha \\
\beta-\gamma\\
\end{array}\hspace*{-1mm}
\right) \bigg\{ \zeta\e^{2\upsilon}\Xi(\chi_{P'})\sinh(\chi_{P'}) \nonumber\\
\espace \ \ -\e^{\upsilon}\left[ (\zeta^2-\lambda^2)\tr \Pi(t')+(\zeta^2+\lambda^2)\tr\Pi^{\dagger}(t') \right]\sinh  (2\chi_{P'}) \nonumber\\
\espace \ \ +\left[ \lambda^2/\zeta\cdot((\tr\Pi^{\dagger}(t'))^2-|\tr \Pi(t')|^2)+4\zeta\det \Pi(t') +2\zeta(|\tr\Pi(t')|^2-1)\right]\sinh (\chi_{P'}) \bigg\}\nonumber\\
\espace+\mathrm{h.c.},\\
f_y \eq 32\lambda_0^2\lambda\e^{-3\upsilon}\textrm{Im}\{\alpha^{\ast}(\beta-\gamma)\}\Xi^{-2}(\chi_{P'}) \nonumber\\
\espace \ \ \times \left[ 2\zeta\e^\upsilon \sinh (2\chi_{P'})-(\mathrm{tr}\Pi(t')+\textrm{tr}\Pi^{\dagger}(t'))\sinh(\chi_{P'})\right],
\end{eqnarray}
where $\Xi(\chi_{P'})$ is an even function of $\chi_{P'}$, $\chi_{P'}(x,t)$ is a parallel-shifted coordinate, and $\upsilon$ is a constant:
\begin{eqnarray}
\Xi(\chi_{P'})\eequiv2\cosh(2\chi_{P'})-\frac{2\e^{-\upsilon}}{\xi}(\tr \Pi(t')+\tr\Pi^{\dagger}(t'))\cosh(\chi_{P'})\nonumber\\
\espace+\frac{\zeta^2+|\tr \Pi(t')|^2}{\lambda^2\det\Pi(t')}+\frac{2}{\lambda^2},\\
\chi_{P'}\eequiv \chi_{P}+\upsilon=-2\lambda_0\zeta(x-2kt')+\upsilon, \\
\upsilon\eequiv \ln \left[\frac{\lambda}{\zeta}(\textrm{det}\Pi(t'))^{1/2}\right].
\end{eqnarray}
Note that $f_x$ and $f_z$ share the same functional form. Each component of the above local spin density is an odd function of $\chi_{P'}$ and, in particular, it has the node at the same point $x_0$ such that $\chi_{P'}(x_0,t')=0$, namely, $x_0=2kt'+(2\lambda_0\zeta)^{-1}\upsilon$. Consequently, the total spin amounts to zero:
\begin{equation}
\vecvar{F}_\mathrm{T}=\int \d x \vecvar{f}(\chi_{P'})=(0,0,0)^{T}.
\end{equation}
For this reason, the PS-type of solitons, on the average, belongs to the polar state \cite{Ho}. 

\section{Two-Soliton Collision}
\label{sec:two-soliton}

We proceed to the discussion of two-soliton collisions in the integrable spinor model (\ref{stdeq}). Two-soliton solutions are obtained by setting $N=4$ ($M=2$) in the formula (\ref{Nsoliton}). There exist two independent discrete eigenvalues and symmetric polarization matrices, respectively, i.e., $\lambda_1=\lambda_2^*$ and $\Pi_1=\Pi_2^{\dagger}$ for one of solitons, $\lambda_3=\lambda_4^*$ and $\Pi_3=\Pi_4^{\dagger}$ for the other. Each soliton is separated at $t=\pm \infty$. Then, a two-soliton is asymptotically two one-solitons. The classification of one-soliton solutions based on the values of the determinants of polarization matrices, discussed in the previous section, is thus valid for two-soliton solutions. 
 
The derivation of the explicit form is more complicated than that in the case for one-soliton solutions. For the derivative NLS equation under NVBC, explicit two-soliton solutions and shifts of soliton positions due to collisions between solitons have been analytically obtained, in the case of purely imaginary eigenvalues, where complexity of calculation is considerably reduced \cite{dNLS}. This strategy, however, does not stand in our NLS equation under NVBC. The reason is understood from eq. (\ref{PNVBC11}). In the spinor model, purely imaginary eigenvalues give two solitons with the same velocity $2k$, and they do not collide with each other. Accordingly, we can not investigate the properties of collisions for purely imaginary discrete eigenvalues. No one has studied explicit multi-soliton solutions of the NLS equation under NVBC, even in the single-component case, because of the computational complexity.

Here, we graphically show the characteristic behaviors of two-soliton collisions in the spinor model, by use of the exact solutions given by the ISM. Referring to them, we carry out the qualitative discussions. Although the presented graphs are depicted for the specific parameters, much the same behaviors are observed for arbitrarily selected parameters. 

\begin{figure}[htbp]
\begin{center}
\begin{picture}(370,110)(0,-5)
\put(56,-14){(a)}
\put(169,-14){(b)}
\put(283,-14){(c)}
\includegraphics[width=\linewidth]{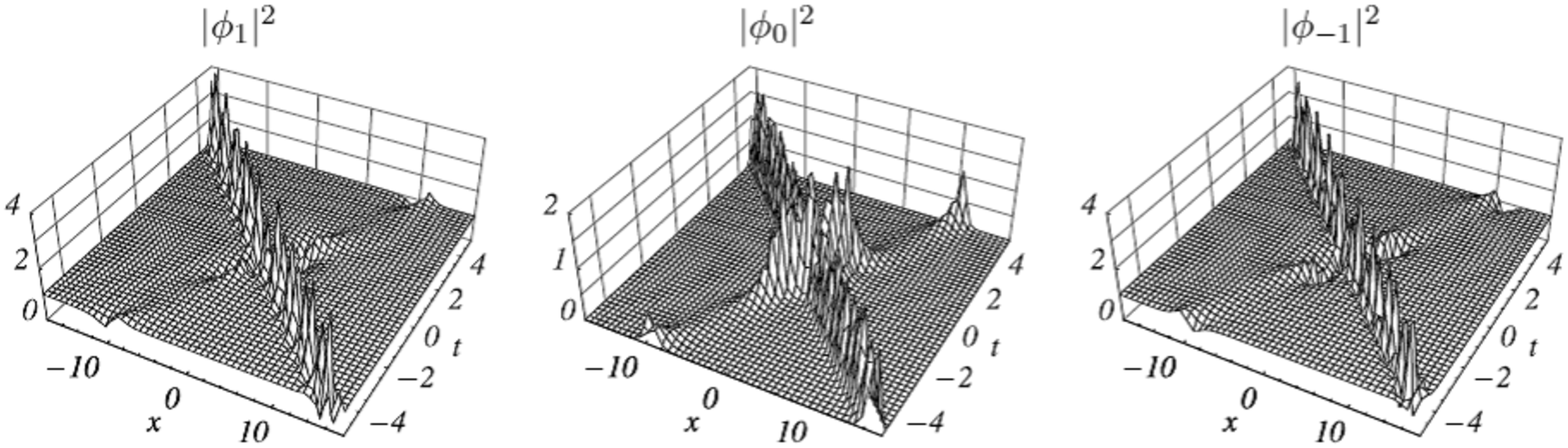}
\end{picture}
\caption{Density plots of $|\phi_1|^2$ (a), $|\phi_0|^2$ (b) and $|\phi_{-1}|^2$ (c) for a mutual collision between two PS-types. The parameters used here are $k=1$, $\lambda_0=1$, $\lambda_1=1.03\im$, $\lambda_3=1.05+\im$, $\Pi_1=\left(\hspace*{-1mm}\begin{array}{cc}
1/\sqrt{2} \hspace*{-1mm}&\hspace*{-1mm}\im/2\\
\im/2\hspace*{-1mm}&\hspace*{-1mm}0
\end{array}\hspace*{-1mm}\right)$, $\Pi_3=\left(\hspace*{-1mm}\begin{array}{cc}
0 \hspace*{-1mm}&\hspace*{-1mm}\im/2\\
\im/2\hspace*{-1mm}&\hspace*{-1mm}1/\sqrt{2}
\end{array}\hspace*{-1mm}\right)$. The velocity of the right (left) moving soliton is $2.00$ ($-3.41$). The collision takes place at $t=0$.
}
\label{fig:ppcollision}
\end{center}
\end{figure}

Figure \ref{fig:ppcollision} illustrates the behavior of a mutual collision between two PS-types, where $\det \Pi_1\neq 0$ and $\det \Pi_3\neq 0$. One can see that, in all three components, Both solitons retain their shapes before and after the collision, which is the common property with solitons in the single-component case. In this sense, PS-PS soliton collision is essentially equivalent to two-soliton collision of the single-component NLS equation.  

\begin{figure}[htbp]
\begin{center}
\begin{picture}(370,110)(0,-5)
\put(56,-14){(a)}
\put(169,-14){(b)}
\put(283,-14){(c)}
\includegraphics[width=\linewidth]{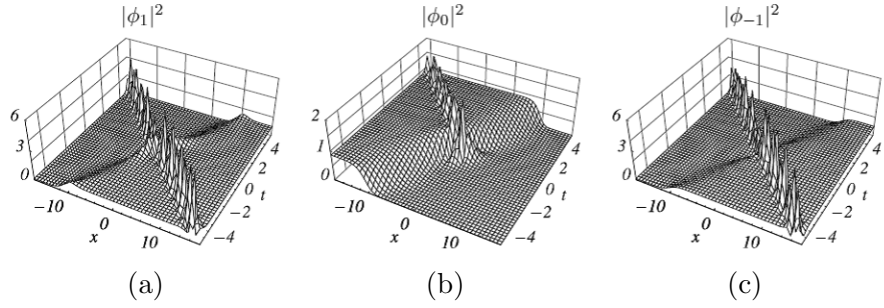}
\end{picture}
\caption{Density plots of $|\phi_1|^2$ (a), $|\phi_0|^2$ (b) and $|\phi_{-1}|^2$ (c) for a mutual collision between DW-type and PS-type. The parameters used here are the same as those of Fig. \ref{fig:ppcollision}, except for $\Pi_1=\left(\hspace*{-1mm}\begin{array}{cc}
2/3 \hspace*{-1mm}&\hspace*{-1mm}\sqrt{2}\im/3\\
\sqrt{2}\im/3\hspace*{-1mm}&\hspace*{-1mm}-1/3
\end{array}\hspace*{-1mm}\right)$, $\Pi_3=\left(\hspace*{-1mm}\begin{array}{cc}
1/\sqrt{2} \hspace*{-1mm}&\hspace*{-1mm}0\\
0\hspace*{-1mm}&\hspace*{-1mm}-1/\sqrt{2}
\end{array}\hspace*{-1mm}\right)$. The right (left) moving soliton is DW-type (PS-type).
}
\label{fig:pdcollision}
\end{center}
\end{figure}

Figure \ref{fig:pdcollision} illustrates the behavior of a mutual collision between DW-type and PS-type, where $\det \Pi_1=0$ and $\det \Pi_3\neq 0$. The behavior of collisions between DW-type and PS-type is qualitatively alien from that between two PS-types. One observes that, in PS-type, much of the population initially inhabiting the hyperfine substate $|F=1, m_F=\pm 1\rangle$ is transferred into the hyperfine substate $|F=1, m_F=0\rangle$ due to the collision. In contrast, in DW-type, such spin transfer does not occur, and the domain-wall shape is preserved against the collision. This phenomenon can be interpreted as follows. DW-type, with nonzero spin, can affect the internal spin state of PS-type, whereas PS-type, which is expected to have zero spin in total, does not affect the internal spin state of DW-type. This kind of spin-transfer phenomenon, called the \textit{spin-switching}, has been first predicted for the case of VBC \cite{IMW2}. Due to the conservation laws, the total number, the total spin and so on are invariant throughout the collision process. Population-mixing among internal degrees of freedom is permitted, as far as the conservation laws are not violated. The spin-switching is one of the dynamical processes which make the spinor solitons more interesting.

\begin{figure}[htbp]
\unitlength=1mm
\begin{center}
\begin{picture}(100,70)(0,-7)
\put(23,-9){(a)}
\put(51,-9){(b)}
\put(80,-9){(c)}
\put(23,61){$|\phi_1|^2$}
\put(51,61){$|\phi_0|^2$}
\put(78,61){$|\phi_{-1}|^2$}
\put(12,56){\small{$t$}}
\put(40,56){\small{$t$}}
\put(69,56){\small{$t$}}
\put(33,-4){\small{$x$}}
\put(62,-4){\small{$x$}}
\put(90,-4){\small{$x$}}
\includegraphics[width=.6\linewidth]{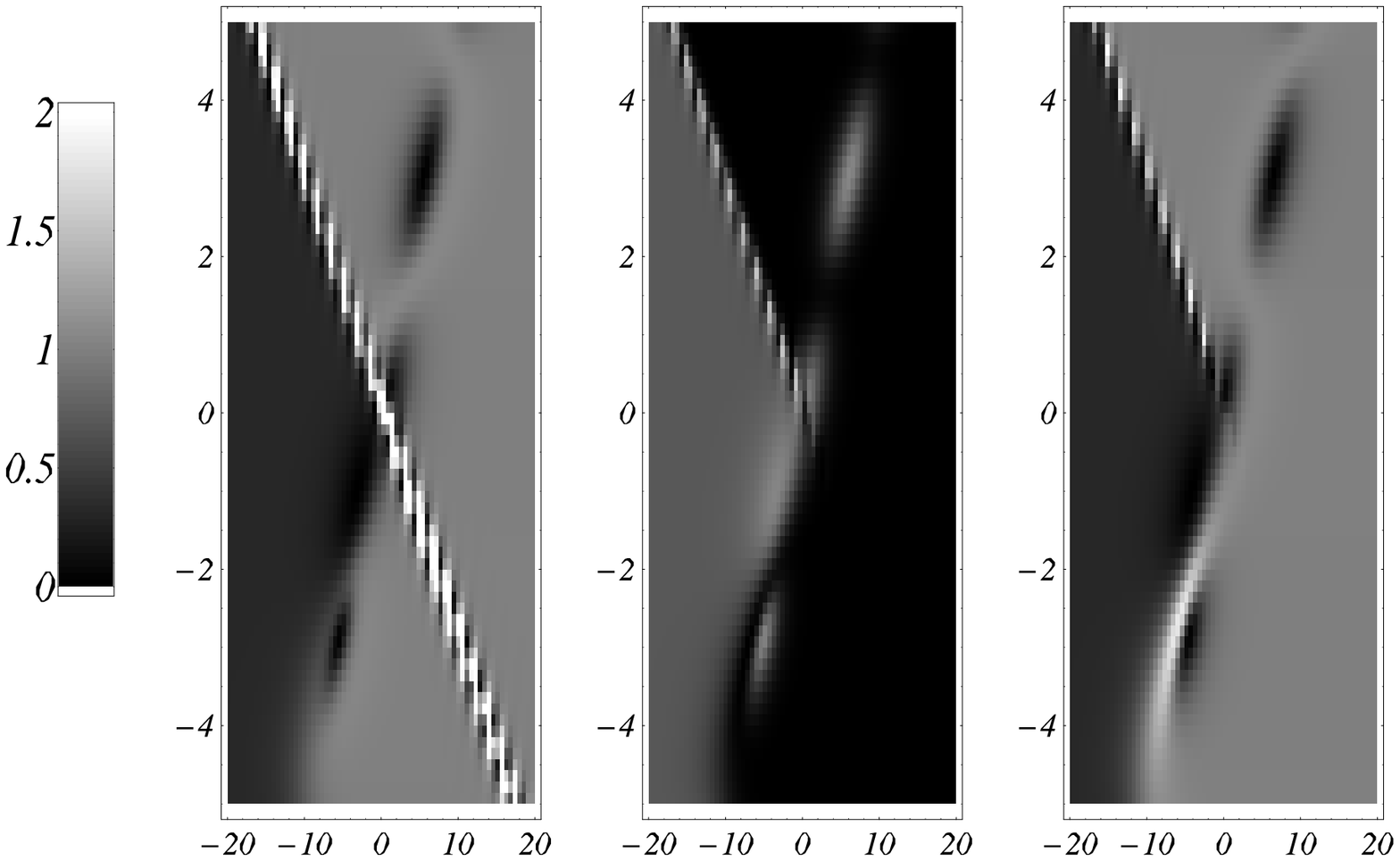}
\end{picture}
\caption{Density plots of $|\phi_1|^2$ (a), $|\phi_0|^2$ (b) and $|\phi_{-1}|^2$ (c) for a mutual collision between two DW-types. The parameters used here are the same as those of Fig. \ref{fig:ppcollision}, except for $\Pi_1=\left(\hspace*{-1mm}\begin{array}{cc}
1/2 \hspace*{-1mm}&\hspace*{-1mm}\im/2\\
\im/2\hspace*{-1mm}&\hspace*{-1mm}-1/2
\end{array}\hspace*{-1mm}\right)$, $\Pi_3=\left(\hspace*{-1mm}\begin{array}{cc}
1 \hspace*{-1mm}&\hspace*{-1mm}0\\
0\hspace*{-1mm}&\hspace*{-1mm}0
\end{array}\hspace*{-1mm}\right)$. The values more than 2 are colored white.
}
\label{fig:ddcollision}
\end{center}
\end{figure}

Finally, for a mutual collision between two DW-types, where $\det \Pi_1=\det \Pi_3= 0$, the shapes of both solitons are expected to be deformed by the collision since each soliton carries nonzero total spin. In fact, drastic population-mixing is seen in  Fig. \ref{fig:ddcollision}, which shows an example of this kind of collisions. One finds that domain-walls "repel" at the collision region, rather than collide.   

\section{Concluding Remarks}
\label{sec:conclusion}

In this paper, we have investigated dynamical properties of matter-wave bright solitons with a finite background in the $F=1$ spinor Bose-Einstein condensate. To perform our analysis concretely, we have exploited an integrable spinor model with a self-focusing nonlinearity and the inverse scattering method under nonvanishing boundary conditions. The situation that matter-wave solitons are located on a finite background fits to the experiments.

One-soliton solutions are derived explicitly and studied in detail. From the point of the mathematical view, they offer general forms of bright soliton solutions of the NLS equation. We have confirmed that our one-soliton solutions include those obtained in the previous works. One-soliton solutions are classified into two kinds by the difference of boundary conditions; DW-type and PS-type. The spin density profiles of one-solitons vary depending on the boundary conditions. In the case of purely imaginary discrete eigenvalues, we have analytically shown that DW-type is in the ferromagnetic state, whereas PS-type is in the polar state. The existence of two distinct magnetic properties for one-soliton solutions also gives rise to fascinating phenomena in the case for two-soliton collisions, for example, the spin-switching. The above results for bright solitons with a finite background agree with those for bright solitons under VBC \cite{IMW1, IMW2}  and dark solitons \cite{UIW}.

Several problems still remain. It is desirable to extend our analysis to the case of general discrete eigenvalues. The computations of the conserved quantities other than the total spin are also required. (One approach is given in Appendix.) In addition, we wish to investigate analytical properties of general $N$-soliton solutions under NVBC in the spinor model. Needless to say, too complicated calculations are inevitable for the above problems. The remaining problems should be discussed elsewhere.
 
We conclude that the properties of the multiple matter-wave solitons in the spinor BECs are interesting and should be useful in various applications. Bright solitons are preferable to dark solitons for applications, because of the advantage in the propagation distance. We hope that our analysis contributes to illuminating dynamical properties of solitons in the spinor BECs, which should be demonstrated experimentally in near future. 

\section*{Acknowledgment}

One of the authors (TK) acknowledges Dr. J. Ieda and Dr. M. Uchiyama for valuable comments and discussions.

\appendix
\section{Several Conserved Quantities of One-Soliton Solutions}

The conserved quantities help us to understand the dynamics of the system. In this appendix, we calculate the total number, the total spin, the total momentum and the total energy of the spinor model. We assume that, in addition to purely imaginary discrete eigenvalues, $\Pi$ is a real symmetric $2\times 2$ matrix. The condition that $\Pi$ is a real symmetric matrix is inherent in the self-defocusing case, i.e., dark solitons \cite{ISM_NVBC}.

For $\Pi=\Pi^{\dagger}$, one-soliton solutions of purely imaginary discrete eigenvalues (\ref{PNVBC1})-(\ref{PNVBC3}) become the following form at $t=t'=(4n-1)\pi /8\lambda_0^2\xi\lambda$ for $n=0,\pm 1,\dots$:
\begin{equation}
\label{realQ}
Q=\lambda_0 \e^{\smalli \phi(x,t)}\left( I +4\im \zeta \frac{\Pi\e^{-(\chi_P+\rho'/2)}+(\sigma^{y}\Pi\sigma^y)\e^{\chi_P+\rho'/2}\det\Pi}{\e^{-(2\chi_P+\rho')}+1+\e^{2\chi_P+\rho'}(\det\Pi)^2}\right),
\end{equation}
where $\e^{\rho'/2}\equiv \lambda/\zeta$. This form is suitable for calculations, since the imaginary part is separated from the real one. One can see clearly that the one-soliton solutions under VBC (\ref{VBC1}) are located on a finite background in the form (\ref{realQ}). Note that the domain-wall shape is lost even for $\det \Pi=0$ there.

Several conserved quantities of the solitons (\ref{realQ}) are calculated by use of eqs. (\ref{tnumber})-(\ref{tenergy}). The results for $\det \Pi=0$ are given by
\begin{eqnarray}
\bar{N}_{\mathrm{T}}\eq 4\lambda_0\zeta, \\
\vecvar{F}_{\mathrm{T}}\eq \bar{N}_{\mathrm{T}}\left(
\begin{array}{c}
2\alpha(\beta+\gamma) \\
0 \\
\beta^2-\gamma^2
\end{array}
\right), \hspace{0.5cm} |\vecvar{F}_{\mathrm{T}}|=\bar{N}_{\mathrm{T}}, \\
\bar{P}_{\mathrm{T}}\eq \bar{N}_{\mathrm{T}}\hbar k, \\
\bar{E}_{\mathrm{T}}\eq \bar{N}_{\mathrm{T}}c\left\{ (k^2-2\lambda_0^2)-\frac{4}{3}\lambda_0^2\zeta^2 \right\},
\end{eqnarray}
and those for $\det \Pi \neq 0$ are given by
\begin{eqnarray}
\bar{N}_{\mathrm{T}}\eq 8\lambda_0\zeta, \\
\vecvar{F}_{\mathrm{T}}\eq (0,0,0)^T, \\
\bar{P}_{\mathrm{T}}\eq \bar{N}_{\mathrm{T}}\hbar k, \\
\bar{E}_{\mathrm{T}}\eq \bar{N}_{\mathrm{T}}c\left\{ (k^2-2\lambda_0^2)-\frac{4}{3}\lambda_0^2\zeta^2 \right\}.
\end{eqnarray}

It is intriguing that, for fixed amplitude and discrete eigenvalue, $\bar{N}_{\mathrm{T}}$, $\bar{P}_{\mathrm{T}}$ and $\bar{E}_{\mathrm{T}}$ of the PS-type ($\det \Pi\neq 0$) have just twice values as 
those of the DW-type ($\det \Pi= 0$), respectively. This enables us to interpret that the PS-type of solitons is a bound state of the two DW-types of solitons. Additionally, for fixed amplitude and total number, the total energy $\bar{E}_{\mathrm{T}}$ of the DW-type is lower than that of the PS-type: $\bar{E}_{\mathrm{T}}^\mathrm{DW}-\bar{E}_{\mathrm{T}}^\mathrm{PS}=-\bar{N}_{\mathrm{T}}^3c/16<0$, which suggests that the DW-type is physically preferable. This result is consistent with inter-atomic ferromagnetic interaction, i.e., $\bar{c}_2<0$.


\begin{thebibliography}{99}
\bibitem{revise}
In ref. \citen{KI78}, the right-hand side of eq. (\ref{KW_sol}) is $\lambda_0+2\lambda_0\zeta\cdots$. We are afraid that there exists a misprint of the sign.
\end{thebibliography}
\end{document}